\documentclass[aps,prd,preprint,superscriptaddress,longbibliography,nofootinbib,floatfix]{revtex4-2}
\usepackage{CJKutf8}
\usepackage{booktabs}
\usepackage{graphicx}
\usepackage{amsmath}
\usepackage{amssymb}
\usepackage{orcidlink}

\begin{document}

\title{Quasinormal-mode residues and pole coalescence
in hypergeometric black-hole models}

\author{Ye Zhou (\begin{CJK*}{UTF8}{gbsn}周烨\end{CJK*})\orcidlink{0009-0004-7050-7736}}
\email{ye.zhou.horizon@gmail.com}
\affiliation{Independent Researcher, Kunshan, Jiangsu, China}

\begin{abstract}
We study source-normalized quasinormal-mode residues in radial boundary-value problems reducible to the Gauss hypergeometric equation. Hypergeometric connection coefficients separate the spectral condition from the normalization and spatial dependence of the frequency-domain Green function. We illustrate this construction for Dirichlet modes of the rotating Ba\~nados--Teitelboim--Zanelli black hole and mixed boundary conditions in two-dimensional anti-de Sitter spacetime, and then apply it to the P\"oschl--Teller/Nariai problem. Near pole coalescence, we obtain finite-position residues for the two simple poles, while at the critical point both coefficients of the second-order Laurent expansion are reduced to closed finite sums. The subleading coefficient depends explicitly on the frequency derivative of the Green-function numerator and is therefore not determined by the spectral function alone. We also identify the range in which the usual vanishing-discriminant criterion characterizes the P\"oschl--Teller coalescence: it holds for the positive barrier considered here, whereas other double zeros or cancellations can occur outside that regime. These results provide a normalization-controlled analytic benchmark for relating spectral coalescence to sourced quasinormal response.
\end{abstract}

\maketitle

\section{Introduction}\label{Sec:Intro}

Quasinormal modes (QNMs) are naturally identified with poles of the frequency-domain Green function of an open gravitational system \cite{Leaver1985,Leaver1986,BertiCardosoStarinets2009,KonoplyaZhidenko2011}. The pole locations determine the complex QNM frequencies, but the strength of the response is encoded in the corresponding residues. This distinction underlies the long-standing study of excitation coefficients and excitation factors \cite{SunPrice1988,NollertPrice1999,BertiCardoso2006,ZhangBertiCardoso2013,Oshita2021}. Recent high-precision calculations have renewed interest in these quantities, particularly near resonant features and avoided crossings in black-hole spectra \cite{Motohashi2025,LoSabaniCardoso2025,OshitaCardoso2025}.

Exactly solvable models provide a useful setting in which the spectrum and the Green function can be treated within the same analytic calculation. Hypergeometric reductions occur in several familiar examples, including BTZ black holes and P\"oschl--Teller-type near-extremal geometries \cite{FerrariMashhoon1984,Birmingham2001,BirminghamSachsSolodukhin2002,CardosoLemos2003}. They also provide a convenient language for asymptotically AdS problems, where the physical boundary condition may involve a linear combination of the two asymptotic branches rather than the elimination of one branch alone \cite{KlebanovWitten1999,Witten2001,IshibashiWald2004,KinoshitaKozukaMurataSugawara2024}. The connection coefficients between endpoint-adapted hypergeometric solutions then contain both the spectral information and the data needed to construct the Green function.

A related issue arises when two QNM poles approach one another. Exceptional points, avoided crossings, and resonantly enhanced excitation have recently been studied in a growing range of black-hole systems \cite{CavalcanteRichartzCarneiroDaCunha2024,Motohashi2025,OshitaBertiCardoso2025,YangBertiFranchini2025,PanossoMacedoKatagiriKubotaMotohashi2026}. In the Nariai limit, the P\"oschl--Teller problem gives a particularly transparent example in which two nearly degenerate poles acquire large individual residues while their combined response remains finite \cite{NakamotoOshita2026}. The underlying higher-order-pole and Jordan structure is known from earlier work on open wave systems and the P\"oschl--Teller potential \cite{Beyer1999,MaassenVanDenBrinkYoung2001}, while a recent Riesz--Laurent formulation gives a general description of sourced response at exceptional points \cite{MorikawaOgawaHirose2026}.

For residue calculations, however, the spectral condition is only part of the problem. A derivative of a quantization function does not by itself fix the residue of a source-normalized Green kernel: the numerator contains the spatial mode profiles and their frequency dependence, and the overall normalization depends on how the delta source is defined. Moreover, singularities present in individual solutions or connection coefficients can cancel in the complete Green function \cite{MotohashiSuichi2026,KubotaMotohashiPoleSkipping2026}. These points become especially relevant near pole coalescence, where the separate simple-pole residues may diverge although the combined response is regular.

In this paper we use hypergeometric connection data to keep these ingredients separate throughout the calculation. The BTZ and AdS$_2$ examples fix the treatment of Dirichlet and mixed boundary conditions and of the associated source normalization. The main explicit application is the P\"oschl--Teller/Nariai problem, for which we evaluate the finite-position residues of the coalescing simple poles and both coefficients of the critical Laurent expansion in closed form. We also clarify the range of validity of the usual discriminant criterion for the P\"oschl--Teller degeneracy. The analysis is restricted to the non-resonant hypergeometric sector; logarithmic endpoint degeneracies are left for separate treatment.

Section~\ref{Sec:Connection} develops the connection and Green-function conventions, Sec.~\ref{Sec:Coalescence} summarizes the local pole-coalescence structure, Sec.~\ref{Sec:Benchmarks} treats the BTZ and AdS$_2$ examples, and Sec.~\ref{Sec:PT} contains the P\"oschl--Teller/Nariai calculation. We conclude in Sec.~\ref{Sec:Conclusion}.

\section{Hypergeometric connection problem}\label{Sec:Connection}
This section fixes the conventions used throughout the paper and collects the part of the calculation common to all three examples. The discussion is local in the complex frequency plane: analyticity is required only in a neighborhood of the pole under consideration. This is sufficient for the residue and Laurent expansions below and avoids implicitly continuing the standard hypergeometric basis through resonant parameter values at which that basis ceases to be regular.

\subsection{Boundary data and spectral condition}\label{SubSec:Boundary}
Consider a radial equation on a compactified interval $0<z<1$,
\begin{equation}
R''(z;\omega)+P(z;\omega)R'(z;\omega)+Q(z;\omega)R(z;\omega)=0,
\label{eq:radial}
\end{equation}
where primes denote derivatives with respect to $z$. We assume that the physical endpoints $z=0$ and $z=1$ are regular singular points. Near $z=0$, write $P=P_0/z+O(1)$ and $Q=Q_0/z^2+O(z^{-1})$. The Frobenius ansatz
\begin{equation}
R(z;\omega)=z^{\rho_0}\sum_{k=0}^{\infty}c_k z^k,\qquad c_0\neq0,
\label{eq:frobenius}
\end{equation}
gives the indicial equation
\begin{equation}
\rho_0(\rho_0-1)+P_0\rho_0+Q_0=0.
\label{eq:indicial}
\end{equation}
The corresponding construction at $z=1$, with $1-z$ as the local variable, defines the exponent $\rho_1$. The physical branch is selected by the boundary condition appropriate to the problem: ingoing or outgoing behavior at a horizon, or a prescribed falloff at a timelike boundary. Radiative QNM conditions are imposed by analytic continuation of the appropriate local wave branch; square integrability at a complex QNM frequency is not assumed.

We factor out the selected endpoint behavior according to
\begin{equation}
R(z;\omega)=z^{\rho_0}(1-z)^{\rho_1}f(z;\omega).
\label{eq:factor}
\end{equation}
The indicial equations remove the leading endpoint singularities. We restrict attention to models for which the residual equation is exactly the Gauss hypergeometric equation,
\begin{equation}
z(1-z)f''+[c-(a+b+1)z]f'-abf=0,
\label{eq:hg}
\end{equation}
with $a=a(\omega)$, $b=b(\omega)$, and $c=c(\omega)$. Regular singular endpoints alone do not imply the reduction \eqref{eq:hg}; exact hypergeometric reducibility is an additional property of the models considered here and will be verified explicitly in each application.

Let $d:=c-a-b$. In the non-resonant chart used below, the solution normalized at the inner endpoint is
\begin{equation}
{R_{\mathrm{in}}}(z;\omega)=z^{\rho_0}(1-z)^{\rho_1}{}_2F_1\!\left(a,b;c;z\right).
\label{eq:Rin}
\end{equation}
A basis adapted to the outer endpoint is
\begin{align}
{R_{\mathrm{s}}}(z;\omega)&=z^{\rho_0}(1-z)^{\rho_1}{}_2F_1\!\left(a,b;1-d;1-z\right),
\label{eq:Rs}\\
{R_{\mathrm{f}}}(z;\omega)&=z^{\rho_0}(1-z)^{\rho_1+d}{}_2F_1\!\left(c-a,c-b;1+d;1-z\right).
\label{eq:Rf}
\end{align}
The subscripts ${\rm s}$ and ${\rm f}$ distinguish the two endpoint branches. They coincide with slow and fast falloff when the real parts of the exponents have that ordering; at a radiative outer endpoint they instead denote the two wave branches, whose incoming or outgoing character will be identified from the local coordinate behavior.

For $d\notin\mathbb Z$, the Kummer connection formula gives \cite{DLMF,Slater1966}
\begin{equation}
\begin{aligned}
{}_2F_1\!\left(a,b;c;z\right)={}&\frac{\Gamma(c)\Gamma(d)}{\Gamma(c-a)\Gamma(c-b)}{}_2F_1\!\left(a,b;1-d;1-z\right)\\
&+\frac{\Gamma(c)\Gamma(-d)}{\Gamma(a)\Gamma(b)}(1-z)^d{}_2F_1\!\left(c-a,c-b;1+d;1-z\right).
\end{aligned}
\label{eq:kummer}
\end{equation}
It follows that
\begin{equation}
{R_{\mathrm{in}}}={C_{\mathrm{s}}}{R_{\mathrm{s}}}+{C_{\mathrm{f}}}{R_{\mathrm{f}}},
\label{eq:connection}
\end{equation}
where
\begin{align}
{C_{\mathrm{s}}}(\omega)&=\frac{\Gamma(c)\Gamma(c-a-b)}{\Gamma(c-a)\Gamma(c-b)},
\label{eq:Cs}\\
{C_{\mathrm{f}}}(\omega)&=\frac{\Gamma(c)\Gamma(a+b-c)}{\Gamma(a)\Gamma(b)}.
\label{eq:Cf}
\end{align}
Once the normalization of \eqref{eq:Rin} has been fixed, these connection coefficients contain the information needed to impose the outer boundary condition.

If a solution is expanded near $z=1$ as $R=c_{\rm s}{R_{\mathrm{s}}}+c_{\rm f}{R_{\mathrm{f}}}$, a linear outer boundary condition can be represented by a nonzero pair $(\alpha,\beta)$ through
\begin{equation}
B_{\alpha,\beta}[R]:=\alpha c_{\rm s}+\beta c_{\rm f}=0.
\label{eq:boundaryfunctional}
\end{equation}
Applying this functional to \eqref{eq:connection} gives the spectral condition
\begin{equation}
F_{\alpha,\beta}(\omega):=\alpha{C_{\mathrm{s}}}(\omega)+\beta{C_{\mathrm{f}}}(\omega)=0.
\label{eq:F}
\end{equation}
Thus the same connection data accommodate pure and mixed boundary conditions without changing the inner normalization. The pair $(\alpha,\beta)$ is projective: multiplying it by a nonzero constant describes the same boundary condition and rescales $F_{\alpha,\beta}$ without changing its zeros.

The formulas above are valid on a local analytic chart, rather than at arbitrary hypergeometric parameter values. In the frequency neighborhood of interest we assume that the selected solution \eqref{eq:Rin} is analytic and nontrivial, the basis \eqref{eq:Rs}--\eqref{eq:Rf} is analytic and linearly independent, and the endpoint powers are defined with fixed branches. A sufficient chart for the unregularized expressions used here is
\begin{equation}
\begin{gathered}
a,b,c,\rho_0,\rho_1\ \text{analytic in }\omega,\\
c\notin\mathbb Z_{\leq0},\qquad d=c-a-b\notin\mathbb Z.
\end{gathered}
\label{eq:analytic-chart}
\end{equation}
The condition on $c$ excludes parameter singularities of the inner-normalized hypergeometric solution, while $d\notin\mathbb Z$ keeps the two Kummer branches independent in the form written above. At integer $d$, logarithmic Frobenius terms can occur and the boundary basis must in general be regularized; special truncated cases can remove the logarithm, so an integer exponent difference by itself does not determine the resonant structure \cite{DLMF}. We therefore do not use Eqs.~\eqref{eq:kummer}--\eqref{eq:F} as an unregularized basis at such points. When an external control parameter is introduced, the required analyticity is that of the physical spectral function and the Green-function numerator. Individual labels $a$ and $b$ may interchange on a square-root cover provided the symmetric combinations entering the physical solution remain analytic.

\subsection{Green function and simple-pole residues}\label{SubSec:Residues}
We next fix the Green-function normalization before discussing residues. Starting from the inhomogeneous equation
\begin{equation}
R''+PR'+QR=J,
\label{eq:inhomogeneous}
\end{equation}
introduce an integrating factor $w$ satisfying
\begin{equation}
\frac{\partial_z w}{w}=P.
\label{eq:wdef}
\end{equation}
The equation then takes the divergence form
\begin{equation}
(wR')'+wQR=wJ.
\label{eq:divergence}
\end{equation}
The differential equation determines $w$ only up to a nonzero factor independent of $z$, which may depend on $\omega$. We keep this normalization distinct from the mode normalization. In each physical application it is fixed by transforming the original inhomogeneous equation together with its unit delta source. This point is essential because rescaling $w$ rescales the Green kernel defined with a unit source in the divergence-form operator.

With the source convention fixed, define
\begin{equation}
\mathcal L_\omega:=\partial_z\!\left(w\partial_z\right)+wQ,
\qquad
\mathcal L_\omega G(z,z';\omega)=\delta(z-z').
\label{eq:Green-definition}
\end{equation}
For a source $J$ in Eq.~\eqref{eq:inhomogeneous}, the corresponding solution is therefore
\begin{equation}
R(z;\omega)=\int_0^1 G(z,z';\omega)w(z';\omega)J(z';\omega)\,\mathrm{d} z'.
\label{eq:Green-inversion}
\end{equation}
This is the standard frequency-domain Green-function construction underlying QNM pole expansions \cite{Leaver1986,NollertSchmidt1992}.

We use the Wronskian convention
\begin{equation}
W[u,v]:=u\,\partial_zv-v\,\partial_zu.
\label{eq:Wronskian}
\end{equation}
A homogeneous solution satisfying the outer boundary condition \eqref{eq:boundaryfunctional} is
\begin{equation}
{R_B}:=\beta{R_{\mathrm{s}}}-\alpha{R_{\mathrm{f}}}.
\label{eq:RB}
\end{equation}
The Green kernel is then
\begin{equation}
G(z,z';\omega)=\frac{{R_{\mathrm{in}}}(z_<;\omega){R_B}(z_>;\omega)}{w(z';\omega)W[{R_{\mathrm{in}}},{R_B}](z';\omega)}.
\label{eq:G}
\end{equation}
Here $z_<:=\min(z,z')$ and $z_>:=\max(z,z')$. The kernel is continuous at $z=z'$, while integration of \eqref{eq:Green-definition} across the source point gives the jump condition
\begin{equation}
w(z';\omega)\left[\left.\partial_zG\right|_{z=z'+0}-\left.\partial_zG\right|_{z=z'-0}\right]=1.
\label{eq:jump}
\end{equation}
For any two homogeneous solutions, Abel's identity gives
\begin{equation}
\partial_zW[u,v]=-P\,W[u,v],\qquad \partial_z\!\left(wW[u,v]\right)=0,
\label{eq:Abel}
\end{equation}
so the Abel-weighted Wronskian is independent of the radial evaluation point.

Using Eqs.~\eqref{eq:connection} and \eqref{eq:RB}, the Wronskian factorizes as
\begin{equation}
W[{R_{\mathrm{in}}},{R_B}]=-F_{\alpha,\beta}(\omega){W_{\mathrm{sf}}},
\qquad {W_{\mathrm{sf}}}:=W[{R_{\mathrm{s}}},{R_{\mathrm{f}}}].
\label{eq:Wfact}
\end{equation}
Indeed,
\begin{equation}
W[{R_{\mathrm{in}}},{R_B}]=-\alpha{C_{\mathrm{s}}} W[{R_{\mathrm{s}}},{R_{\mathrm{f}}}]+\beta{C_{\mathrm{f}}} W[{R_{\mathrm{f}}},{R_{\mathrm{s}}}]=-(\alpha{C_{\mathrm{s}}}+\beta{C_{\mathrm{f}}}){W_{\mathrm{sf}}}.
\label{eq:Wfact-proof}
\end{equation}
The basis Wronskian has a particularly simple Abel normalization. Reconstructing the equation for $R=z^{\rho_0}(1-z)^{\rho_1}f$ from \eqref{eq:hg}, the coefficient of $R'$ and a corresponding integrating factor are
\begin{align}
P_R(z;\omega)&=\frac{c-2\rho_0}{z}+\frac{d-1+2\rho_1}{1-z},
\label{eq:PR}\\
w_0(z;\omega)&=z^{c-2\rho_0}(1-z)^{1-d-2\rho_1}.
\label{eq:w0}
\end{align}
Since ${R_{\mathrm{s}}}\sim(1-z)^{\rho_1}$ and ${R_{\mathrm{f}}}\sim(1-z)^{\rho_1+d}$ as $z\to1$, one finds
\begin{equation}
{W_{\mathrm{sf}}}\sim-d(1-z)^{2\rho_1+d-1},\qquad z\to1,
\label{eq:Wsf-asymptotic}
\end{equation}
and Abel's identity therefore gives the exact constant
\begin{equation}
w_0{W_{\mathrm{sf}}}=-d.
\label{eq:w0W}
\end{equation}
For the physically normalized integrating factor we may write
\begin{equation}
w(z;\omega)=\mathcal N(\omega)w_0(z;\omega),\qquad D(\omega):=w{W_{\mathrm{sf}}}=-\mathcal N(\omega)d.
\label{eq:D}
\end{equation}
In the local pole calculation we assume that the source-normalization factor $\mathcal N(\omega)$, and hence $D(\omega)$, is analytic and nonzero. Its explicit value will be fixed separately in the BTZ, AdS$_2$, and P\"oschl--Teller problems.

Combining Eqs.~\eqref{eq:G}, \eqref{eq:Wfact}, and \eqref{eq:D} gives the useful quotient form
\begin{equation}
G(z,z';\omega)=\frac{H(z,z';\omega)}{F_{\alpha,\beta}(\omega)},
\qquad
H(z,z';\omega):=-\frac{{R_{\mathrm{in}}}(z_<;\omega){R_B}(z_>;\omega)}{D(\omega)}.
\label{eq:HF}
\end{equation}
This representation separates the spectral zero from the source normalization and the spatial mode profiles.

Let $\omega_n$ be a simple spectral root,
\begin{equation}
F_{\alpha,\beta}(\omega_n)=0,
\qquad
\partial_\omega F_{\alpha,\beta}(\omega_n)\neq0.
\label{eq:simple-root}
\end{equation}
Equation \eqref{eq:Wfact}, together with the nonvanishing of ${W_{\mathrm{sf}}}$, shows that ${R_{\mathrm{in}}}$ and ${R_B}$ are linearly dependent at $\omega_n$. We write
\begin{equation}
{R_B}(z;\omega_n)=\chi_n{R_{\mathrm{in}}}(z;\omega_n).
\label{eq:chi-definition}
\end{equation}
The root condition gives the proportionality factor directly:
\begin{equation}
\chi_n=
\begin{cases}
\displaystyle \frac{\beta}{{C_{\mathrm{s}}}(\omega_n)}, & \beta\neq0,\\[2mm]
\displaystyle -\frac{\alpha}{{C_{\mathrm{f}}}(\omega_n)}, & \alpha\neq0.
\end{cases}
\label{eq:chi}
\end{equation}
When both $\alpha$ and $\beta$ are nonzero the two expressions agree. The denominator relevant to either branch cannot vanish at the root: otherwise Eq.~\eqref{eq:F} would force both ${C_{\mathrm{s}}}$ and ${C_{\mathrm{f}}}$ to vanish, and the exact expansion \eqref{eq:connection} would imply the normalized solution ${R_{\mathrm{in}}}$ to be identically zero.

Expanding \eqref{eq:HF} around the simple zero gives the residue kernel
\begin{equation}
\operatorname*{Res}_{\omega=\omega_n}G(z,z';\omega)=-\frac{\chi_n{R_{\mathrm{in}}}(z;\omega_n){R_{\mathrm{in}}}(z';\omega_n)}{D(\omega_n)\,\partial_\omega F_{\alpha,\beta}(\omega_n)}.
\label{eq:residue}
\end{equation}
This is a statement about the Green kernel at fixed interior points. A nontrivial mode can have nodes, and source or observation projections can likewise suppress the pole in a particular matrix element. The multiplicity of the spectral zero must therefore be distinguished from the value of a chosen projected residue.

The construction is also regular under a change of outer basis. Let
\begin{equation}
\mathbf R:=({R_{\mathrm{s}}},{R_{\mathrm{f}}}),\qquad \mathbf C:=({C_{\mathrm{s}}},{C_{\mathrm{f}}})^T,\qquad \mathbf b:=(\alpha,\beta)^T,
\qquad
\Omega:=\begin{pmatrix}0&1\\-1&0\end{pmatrix}.
\label{eq:vector-notation}
\end{equation}
Then ${R_{\mathrm{in}}}=\mathbf R\mathbf C$, $F=\mathbf b^T\mathbf C$, and ${R_B}=\mathbf R\Omega\mathbf b$. For an analytic invertible matrix $M(\omega)$, define
\begin{equation}
\widetilde{\mathbf R}=\mathbf R M,
\qquad
\widetilde{\mathbf C}=M^{-1}\mathbf C,
\qquad
\widetilde{\mathbf b}=M^T\mathbf b.
\label{eq:basis-change}
\end{equation}
It follows that $\widetilde F=F$. Moreover, using $M\Omega M^T=(\det M)\Omega$,
\begin{equation}
\widetilde{{R_B}}=(\det M){R_B},
\qquad
\widetilde{{W_{\mathrm{sf}}}}=(\det M){W_{\mathrm{sf}}},
\label{eq:basis-covariance}
\end{equation}
so the Green kernel \eqref{eq:G} is unchanged. If $M$ depends on frequency, the induced frequency dependence of $\widetilde{\mathbf b}$ must be retained when differentiating the spectral function; with this dependence included, $\widetilde F(\omega)=F(\omega)$ identically. Likewise, a nonzero analytic rescaling ${R_{\mathrm{in}}}\mapsto s(\omega){R_{\mathrm{in}}}$ multiplies both $H$ and $F$ by $s$ and leaves $G=H/F$, its poles, and their multiplicities unchanged. Singular rescalings are not included in this statement.

For the hypergeometric connection coefficients, the derivatives entering \eqref{eq:residue} can be evaluated algebraically away from Gamma-function poles. Writing $\psi(x)=\Gamma'(x)/\Gamma(x)$ and $a_\omega:=\partial_\omega a$ (and similarly for $b$ and $c$), one obtains
\begin{equation}
\begin{aligned}
\partial_\omega{C_{\mathrm{s}}}={}&{C_{\mathrm{s}}}\Big[c_\omega\psi(c)+(c_\omega-a_\omega-b_\omega)\psi(c-a-b)\\
&\hspace{2.8cm}-(c_\omega-a_\omega)\psi(c-a)-(c_\omega-b_\omega)\psi(c-b)\Big],
\end{aligned}
\label{eq:Csder}
\end{equation}
and
\begin{equation}
\begin{aligned}
\partial_\omega{C_{\mathrm{f}}}={}&{C_{\mathrm{f}}}\Big[c_\omega\psi(c)+(a_\omega+b_\omega-c_\omega)\psi(a+b-c)\\
&\hspace{2.8cm}-a_\omega\psi(a)-b_\omega\psi(b)\Big].
\end{aligned}
\label{eq:Cfder}
\end{equation}
At a zero generated by a reciprocal Gamma function, these product forms should be evaluated by analytic continuation rather than by multiplying a numerical zero by a divergent Digamma function. If $g(x):=1/\Gamma(x)$ and $n\in\mathbb N_0$, then \cite{DLMF}
\begin{equation}
g(-n+t)=(-1)^n n!\,t\left[1-\psi(n+1)t+O(t^2)\right],\qquad t\to0,
\label{eq:rgamma}
\end{equation}
so that
\begin{equation}
g'(-n)=(-1)^n n!.
\label{eq:rgamma-derivative}
\end{equation}
This local expansion will be used repeatedly in the explicit residue calculations.

Equation \eqref{eq:residue} also makes clear why $\partial_\omega F$ by itself is not a normalized excitation amplitude: the full pole coefficient contains the numerator $H$, including the spatial mode profiles and the unit-source normalization. More generally, a pole or zero of an individual local solution or connection coefficient need not survive in the complete Green kernel after all factors are assembled; recent Kerr analyses provide explicit examples of such cancellations \cite{MotohashiSuichi2026,KubotaMotohashiPoleSkipping2026}. The pole-coalescence analysis in the next section is therefore carried out directly at the level of the quotient \eqref{eq:HF}, retaining the frequency dependence of both numerator and denominator.

\section{Local structure near pole coalescence}\label{Sec:Coalescence}
The quotient form obtained in Eq.~\eqref{eq:HF} makes the local coalescence problem largely independent of the details of the radial equation. Pole coalescence and the associated Jordan structure are well known in open and non-Hermitian wave systems \cite{MaassenVanDenBrinkYoung2001,Heiss2012}, and have recently become important in studies of resonant black-hole response \cite{Motohashi2025,NakamotoOshita2026,MorikawaOgawaHirose2026}. We use only the local analytic consequences needed below. In particular, the Laurent identities in this section are not new spectral results; they identify the quantities that will later be evaluated explicitly from the hypergeometric connection data. The spatial arguments $z,z'$ are held fixed and suppressed when no ambiguity can arise.

\subsection{Double-pole Laurent expansion}\label{SubSec:Laurent}
Let $\zeta:=\omega-\omega_*$ and suppose that $F$ and $H$ are analytic near $\omega_*$. For fixed $z,z'$, assume
\begin{equation}
F_*=0,\qquad F_{\omega,*}=0,\qquad F_{\omega\omega,*}\neq0,\qquad H_*\neq0,
\label{eq:double-root}
\end{equation}
where a subscript $*$ denotes evaluation at $\omega_*$. The denominator then has a zero of exactly second order while the numerator remains nonzero. Expanding to the order required for the complete principal part gives
\begin{align}
F(\omega)&=\frac{1}{2}F_{\omega\omega,*}\zeta^2+\frac{1}{6}F_{\omega\omega\omega,*}\zeta^3+O(\zeta^4),
\label{eq:F-double-expansion}\\
H(\omega)&=H_*+H_{\omega,*}\zeta+O(\zeta^2).
\label{eq:H-double-expansion}
\end{align}
Substitution into $G=H/F$ yields
\begin{equation}
G(z,z';\omega)=\frac{K_2(z,z')}{\zeta^2}+\frac{K_1(z,z')}{\zeta}+O(1),
\label{eq:double-Laurent}
\end{equation}
where
\begin{align}
K_2&=\frac{2H_*}{F_{\omega\omega,*}},
\label{eq:K2-general}\\
K_1&=\frac{2H_{\omega,*}}{F_{\omega\omega,*}}-\frac{2H_*F_{\omega\omega\omega,*}}{3F_{\omega\omega,*}^{\,2}}.
\label{eq:K1-general}
\end{align}
The same quotient identities arise in the recent Riesz--Laurent treatment of sourced black-hole response \cite{MorikawaOgawaHirose2026}. The second coefficient is particularly important here: $K_1$ depends not only on the cubic term in the spectral function but also on the frequency derivative of the Green-function numerator. With the normalization of Sec.~\ref{Sec:Connection},
\begin{equation}
H_*=-\frac{\chi_*{R_{\mathrm{in}}}(z;\omega_*){R_{\mathrm{in}}}(z';\omega_*)}{D_*},
\label{eq:Hstar}
\end{equation}
and hence
\begin{equation}
K_2=-\frac{2\chi_*{R_{\mathrm{in}}}(z;\omega_*){R_{\mathrm{in}}}(z';\omega_*)}{D_*F_{\omega\omega,*}}.
\label{eq:K2-normalized}
\end{equation}
The derivative $H_{\omega,*}$ in Eq.~\eqref{eq:K1-general} also contains the frequency dependence of the mode profiles, the boundary-adapted solution, and the source-normalization factor $D$. Differentiating only the spectral denominator would therefore miss part of the simple-pole coefficient.

The condition $H_*\neq0$ is a condition on the particular kernel element under consideration. For the pointwise Green kernel it can fail, for example, when one of the selected positions lies at a node of the critical mode. After spatial projection, an analogous cancellation may occur through the source or observer functional. Such cancellations can reduce the visible pole order without changing the multiplicity of the zero of $F$. Conversely, within the analytic chart of Sec.~\ref{Sec:Connection}, a second-order pole with nonvanishing numerator requires a double zero of the spectral function.

\subsection{Nearby poles and the finite two-mode response}\label{SubSec:Unfolding}
Let $p$ be a control parameter and set $q:=p-p_*$. We assume that $F(\omega,p)$ and $H(\omega,p)$ are jointly analytic near $(\omega_*,p_*)$ and that
\begin{equation}
F_{p,*}\neq0.
\label{eq:transverse}
\end{equation}
This is the generic one-parameter unfolding of the double zero; if $F_{p,*}=0$, the leading splitting can occur at a different order. The analyticity assumption applies to the physical spectral function and Green-function numerator, not necessarily to a particular labeling of the hypergeometric parameters. Individual quantities such as $a$ and $b$ may interchange on a square-root cover while the combinations entering $F$ and $H$ remain analytic.

Keeping the terms that determine the two roots through order $q$, write
\begin{equation}
F=A\zeta^2+Bq+C\zeta^3+E\zeta q+\mathcal R(\zeta,q),
\qquad
\mathcal R=O\!\left((|\zeta|^2+|q|)^2\right),
\label{eq:F-unfold}
\end{equation}
with
\begin{equation}
A:=\frac{1}{2}F_{\omega\omega,*},\qquad
B:=F_{p,*},\qquad
C:=\frac{1}{6}F_{\omega\omega\omega,*},\qquad
E:=F_{\omega p,*}.
\label{eq:ABCE}
\end{equation}
Set $q=\tau^2$ and $\zeta=\tau y$. At $\tau=0$, the reduced equation has the two simple roots $y=\pm s$, where
\begin{equation}
s^2=-\frac{2F_{p,*}}{F_{\omega\omega,*}}.
\label{eq:s-coeff}
\end{equation}
The analytic implicit-function theorem then gives two branches on the local $\tau$-cover,
\begin{equation}
\omega_\pm=\omega_*\pm s\sqrt q+bq+O(q^{3/2}),
\label{eq:root-splitting}
\end{equation}
with
\begin{equation}
b=-\frac{F_{\omega p,*}}{F_{\omega\omega,*}}+\frac{F_{\omega\omega\omega,*}F_{p,*}}{3F_{\omega\omega,*}^{\,2}}.
\label{eq:b-coeff}
\end{equation}
Because the zero of $F(\omega,p_*)$ at $\omega_*$ is isolated and has multiplicity two, Rouch\'e's theorem ensures that these are the only two roots inside a sufficiently small fixed contour for sufficiently small $q$. Changing the branch of $\sqrt q$ merely interchanges their labels.

For $q\neq0$ sufficiently small, the two roots are simple. Their residue kernels are
\begin{equation}
\mathcal R_\pm(z,z'):=\operatorname*{Res}_{\omega=\omega_\pm}G(z,z';\omega)=\frac{H(\omega_\pm,p)}{F_\omega(\omega_\pm,p)}.
\label{eq:Rpm-definition}
\end{equation}
Using Eqs.~\eqref{eq:F-unfold}--\eqref{eq:b-coeff}, one finds
\begin{equation}
F_\omega(\omega_\pm,p)=\pm F_{\omega\omega,*}s\sqrt q+\frac{1}{3}F_{\omega\omega\omega,*}s^2q+O(q^{3/2}),
\label{eq:Fomega-roots}
\end{equation}
while
\begin{equation}
H(\omega_\pm,p)=H_*\pm H_{\omega,*}s\sqrt q+O(q).
\label{eq:H-roots}
\end{equation}
Their ratio therefore gives
\begin{equation}
\mathcal R_\pm(z,z')=\pm\frac{K_2(z,z')}{2s\sqrt q}+\frac{K_1(z,z')}{2}+O(\sqrt q).
\label{eq:residue-unfolding}
\end{equation}
When $K_2\neq0$, the individual residues thus grow as $q^{-1/2}$ with opposite leading complex amplitudes, whereas their finite parts approach the same value. This local behavior is consistent with the enhanced excitation and near-opposite phases found close to exceptional points in recent black-hole studies \cite{Motohashi2025,NakamotoOshita2026,RossiOshitaGualtieriBerti2026}. A source or observation zero can remove the leading divergence from a projected response, and the scaling need not hold for a nongeneric unfolding with $F_{p,*}=0$.

The divergent individual residues are better organized into branch-independent combinations. Define
\begin{align}
\omega_c&:=\frac{\omega_++\omega_-}{2},\qquad
\delta:=\frac{\omega_+-\omega_-}{2},
\label{eq:center-splitting}\\
S&:=\mathcal R_++\mathcal R_-,\qquad
T:=\delta(\mathcal R_+-\mathcal R_-).
\label{eq:ST}
\end{align}
On the $\tau=\sqrt q$ cover, changing $\tau\to-\tau$ exchanges the two roots. The Laurent expansions of $\omega_c$, $\delta^2$, $S$, and $T$ are therefore even in $\tau$. Equation \eqref{eq:residue-unfolding} shows explicitly that the possible negative powers cancel in $S$ and $T$, so all four combinations extend analytically to $\tau=0$ and hence are analytic functions of $q$. Their leading behavior is
\begin{equation}
\omega_c=\omega_*+bq+O(q^2),\qquad
S=K_1+O(q),\qquad
T=K_2+O(q).
\label{eq:ST-limits}
\end{equation}
The principal part associated with the pair can then be written exactly as
\begin{equation}
\frac{\mathcal R_+}{\omega-\omega_+}+\frac{\mathcal R_-}{\omega-\omega_-}
=
\frac{S(\omega-\omega_c)+T}{(\omega-\omega_c)^2-\delta^2}.
\label{eq:pair-principal-part}
\end{equation}
Its limit as $q\to0$ is precisely the double-pole principal part in Eq.~\eqref{eq:double-Laurent}, without requiring the subtraction of two separately divergent residues.

The same cancellation appears directly in the time dependence:
\begin{equation}
\mathcal R_+e^{-\mathrm{i}\omega_+t}+\mathcal R_-e^{-\mathrm{i}\omega_-t}
=
e^{-\mathrm{i}\omega_ct}
\left[
S\cos(\delta t)-\mathrm{i} T\frac{\sin(\delta t)}{\delta}
\right].
\label{eq:pair-time}
\end{equation}
For fixed $t$, the coalescence limit is
\begin{equation}
\mathcal R_+e^{-\mathrm{i}\omega_+t}+\mathcal R_-e^{-\mathrm{i}\omega_-t}
\longrightarrow
e^{-\mathrm{i}\omega_*t}\left(K_1-\mathrm{i} tK_2\right),
\label{eq:pair-time-limit}
\end{equation}
with an $O(q)$ error uniformly on compact time intervals. This limit should not be replaced by a small-$\delta t$ approximation on time scales $t=O(|q|^{-1/2})$, for which the exact expression \eqref{eq:pair-time} must be retained.

For definiteness, we adopt the inverse-frequency convention
\begin{equation}
g(t)=\int_{\mathbb R+\mathrm{i}\gamma}\frac{\mathrm{d}\omega}{2\pi}\,e^{-\mathrm{i}\omega t}G(\omega),
\label{eq:inverse-transform}
\end{equation}
with the contour chosen above the relevant singularities. When the contour is deformed downward for $t>0$, a clockwise enclosure of the isolated pole pair contributes $-\mathrm{i}$ times the corresponding frequency-domain residues. The critical pair contribution is therefore
\begin{equation}
g_{\rm pair}^{\rm crit}(t)=e^{-\mathrm{i}\omega_*t}\left(-\mathrm{i} K_1-tK_2\right).
\label{eq:critical-time-response}
\end{equation}
The linear factor is the familiar time-domain signature of a second-order pole \cite{MaassenVanDenBrinkYoung2001,YangBertiFranchini2025,NakamotoOshita2026}. Equation \eqref{eq:critical-time-response} refers only to this isolated pole cluster. Other poles, branch cuts where present, and non-pole contour contributions may be required for the complete causal response \cite{Beyer1999,Kuntz2026,ArnaudoCarballoWithers2026}.

Finally, the coefficient observed in a sourced problem depends on the source as well as on the Green kernel. Let $\mathcal S(z',\omega)$ denote the source density on the right-hand side of the divergence-form equation and assume
\begin{equation}
\mathcal S(z',\omega)=\mathcal S_*(z')+\zeta\,\mathcal S_{\omega,*}(z')+O(\zeta^2).
\label{eq:source-expansion}
\end{equation}
Provided the spatial integral can be interchanged with the local Laurent expansion, the coefficients of the projected response are
\begin{align}
\mathcal K_2(z)&=\int_0^1 K_2(z,z')\mathcal S_*(z')\,\mathrm{d} z',
\label{eq:projected-K2}\\
\mathcal K_1(z)&=\int_0^1\left[K_1(z,z')\mathcal S_*(z')+K_2(z,z')\mathcal S_{\omega,*}(z')\right]\mathrm{d} z'.
\label{eq:projected-K1}
\end{align}
An additional observer projection can be applied in the same way. Thus the double zero of $F$, the Laurent coefficients of the Green kernel, and a source-dependent excitation amplitude are closely related but distinct quantities. With this local structure fixed, we next turn to the boundary normalization in the BTZ and AdS$_2$ examples before evaluating the coalescing P\"oschl--Teller residues explicitly.

\section{Boundary conditions and normalization in BTZ and AdS$_2$}\label{Sec:Benchmarks}

We now apply the preceding formulas to two AdS examples. The rotating BTZ scalar field provides a useful benchmark because its quasinormal spectrum and Green-function poles are known analytically, whereas the AdS$_2$ problem illustrates how mixed boundary data enter the same connection framework. In both cases the emphasis is on the boundary condition and the normalization of the unit-source Green function.

\subsection{Dirichlet residues in BTZ}\label{SubSec:BTZ}

Set the AdS radius to unity and consider a scalar field of mass $\mu$ on a rotating BTZ black hole with $r_+>r_-\geq0$ \cite{BanadosTeitelboimZanelli1992}. For
\begin{equation}
\Phi=e^{-\mathrm{i}\omega t+\mathrm{i} m\varphi}R(r),\qquad m\in\mathbb Z,
\label{eq:BTZ-ansatz}
\end{equation}
the radial Klein--Gordon equation is
\begin{equation}
\frac{1}{r}\partial_r\!\left(rN^2\partial_rR\right)+\left[\frac{(\omega+mN^\varphi)^2}{N^2}-\frac{m^2}{r^2}-\mu^2\right]R=0,
\label{eq:BTZ-radial}
\end{equation}
where
\begin{equation}
N^2=\frac{(r^2-r_+^2)(r^2-r_-^2)}{r^2},\qquad N^\varphi=-\frac{r_+r_-}{r^2}.
\label{eq:BTZ-metric-functions}
\end{equation}
Introduce
\begin{equation}
z=\frac{r^2-r_+^2}{r^2-r_-^2},\qquad \Delta_B:=r_+^2-r_-^2,
\label{eq:BTZ-z}
\end{equation}
so that the outer horizon and spatial infinity are mapped to $z=0$ and $z=1$, respectively. Choosing
\begin{equation}
\rho_0=-\frac{\mathrm{i}(\omega r_+-mr_-)}{2\Delta_B},\qquad \rho_1=\frac{1-\nu_B}{2},\qquad \nu_B:=\sqrt{1+\mu^2},
\label{eq:BTZ-exponents}
\end{equation}
reduces the residual equation to the hypergeometric form \eqref{eq:hg}, with
\begin{align}
a&=\rho_0+\rho_1+\frac{\mathrm{i}(\omega r_--mr_+)}{2\Delta_B},\notag\\
b&=\rho_0+\rho_1-\frac{\mathrm{i}(\omega r_--mr_+)}{2\Delta_B},\notag\\
c&=1+2\rho_0.
\label{eq:BTZ-abc}
\end{align}
We work in the non-resonant sector $\nu_B\notin\mathbb Z$ and away from additional Gamma-function singularities. Defining $h_+=(1+\nu_B)/2$, the combinations entering the Dirichlet condition simplify to
\begin{align}
c-a&=h_+-\frac{\mathrm{i}(\omega-m)}{2(r_+-r_-)},\notag\\
c-b&=h_+-\frac{\mathrm{i}(\omega+m)}{2(r_++r_-)}.
\label{eq:BTZ-ca-cb}
\end{align}

The standard Dirichlet condition removes the slow asymptotic branch. This statement is more precise than simply requiring $R\to0$, since both branches can decay in part of the AdS mass window. In the notation of Sec.~\ref{Sec:Connection}, $(\alpha,\beta)=(1,0)$ and
\begin{equation}
F_{1,0}(\omega)={C_{\mathrm{s}}}(\omega)=\frac{\Gamma(c)\Gamma(\nu_B)}{\Gamma(c-a)\Gamma(c-b)}.
\label{eq:BTZ-F}
\end{equation}
Provided no simultaneous singular factor cancels the zero, the QNM condition is
\begin{equation}
c-a=-n\qquad\text{or}\qquad c-b=-n,\qquad n\in\mathbb{N}_0.
\label{eq:BTZ-quantization}
\end{equation}
This gives the familiar two families \cite{Birmingham2001,BirminghamSachsSolodukhin2002,LopezOrtegaMataPacheco2018},
\begin{align}
\omega_n^{(a)}&=m-2\mathrm{i}(r_+-r_-)(n+h_+),\notag\\
\omega_n^{(b)}&=-m-2\mathrm{i}(r_++r_-)(n+h_+).
\label{eq:BTZ-frequencies}
\end{align}
At exceptional parameter values where several Gamma factors become singular simultaneously, the local quotient must instead be expanded before assigning the order of the Green-function pole.

For a simple root on the $c-a=-n$ branch, Eq.~\eqref{eq:rgamma} gives
\begin{equation}
\partial_\omega F_{1,0}(\omega_n^{(a)})=\left.\frac{\Gamma(c)\Gamma(\nu_B)}{\Gamma(c-b)}(-1)^n n!\,\partial_\omega(c-a)\right|_{\omega_n^{(a)}}.
\label{eq:BTZ-Fprime-a}
\end{equation}
Similarly,
\begin{equation}
\partial_\omega F_{1,0}(\omega_n^{(b)})=\left.\frac{\Gamma(c)\Gamma(\nu_B)}{\Gamma(c-a)}(-1)^n n!\,\partial_\omega(c-b)\right|_{\omega_n^{(b)}}.
\label{eq:BTZ-Fprime-b}
\end{equation}
For Dirichlet data, $\chi_n=-1/{C_{\mathrm{f}}}(\omega_n)$, and Eq.~\eqref{eq:residue} becomes
\begin{equation}
\operatorname*{Res}_{\omega=\omega_n^{(\sigma)}}G(z,z';\omega)=\frac{\mathcal A_n^{(\sigma)}}{D_{\rm BTZ}}{R_{\mathrm{in}}}(z;\omega_n^{(\sigma)}){R_{\mathrm{in}}}(z';\omega_n^{(\sigma)}),\qquad \sigma=a,b,
\label{eq:BTZ-residue}
\end{equation}
where
\begin{align}
\mathcal A_n^{(a)}&=\left.\frac{\Gamma(a)\Gamma(b)\Gamma(c-b)}{\Gamma(c)^2\Gamma(-\nu_B)\Gamma(\nu_B)}\frac{(-1)^n}{n!\,\partial_\omega(c-a)}\right|_{\omega_n^{(a)}},\notag\\
\mathcal A_n^{(b)}&=\left.\frac{\Gamma(a)\Gamma(b)\Gamma(c-a)}{\Gamma(c)^2\Gamma(-\nu_B)\Gamma(\nu_B)}\frac{(-1)^n}{n!\,\partial_\omega(c-b)}\right|_{\omega_n^{(b)}}.
\label{eq:BTZ-amplitudes}
\end{align}
The remaining Gamma factors are assumed regular at the selected simple root.

It remains to fix the source normalization. Multiplying Eq.~\eqref{eq:BTZ-radial} by $r$ gives a radial operator with principal coefficient
\begin{equation}
p_B(r):=rN^2.
\label{eq:BTZ-p}
\end{equation}
We define its inverse by a unit source $\delta(r-r')$. Under the change of variable \eqref{eq:BTZ-z},
\begin{equation}
p_B(r)\frac{\mathrm{d} z}{\mathrm{d} r}=2\Delta_B z.
\label{eq:BTZ-weight-transform}
\end{equation}
After dividing the transformed equation by $\mathrm{d} z/\mathrm{d} r$, the identity $\delta(r-r')=(\mathrm{d} z/\mathrm{d} r)_{r'}\delta(z-z')$ leaves a unit $\delta(z-z')$ source. The divergence-form weight is therefore
\begin{equation}
w_{\rm BTZ}(z)=2\Delta_B z.
\label{eq:BTZ-weight}
\end{equation}
For the parameters above, Eq.~\eqref{eq:w0} gives $w_0=z$ and $d=\nu_B$, so
\begin{equation}
D_{\rm BTZ}=w_{\rm BTZ}{W_{\mathrm{sf}}}=-2\Delta_B\nu_B.
\label{eq:BTZ-D}
\end{equation}
Equations \eqref{eq:BTZ-residue} and \eqref{eq:BTZ-D} therefore determine the complete radial Green-kernel residue in this unit-source convention, rather than only its spectral factor $1/\partial_\omega F$.

\subsection{Robin data in AdS$_2$}\label{SubSec:AdS2}

We next consider a minimally coupled probe scalar on the AdS$_2$ black-hole metric
\begin{equation}
\mathrm{d} s^2=-(r^2-r_h^2)\mathrm{d} t^2+(r^2-r_h^2)^{-1}\mathrm{d} r^2,
\label{eq:AdS2-metric}
\end{equation}
again with unit AdS radius. This specifies the probe equation used here; other matter couplings in Jackiw--Teitelboim gravity need not lead to the same radial problem \cite{BhattacharjeeSarkarBhattacharyya2021}. For a scalar of mass $\mu_2$,
\begin{equation}
\partial_r\!\left[(r^2-r_h^2)\partial_rR\right]+\left[\frac{\omega^2}{r^2-r_h^2}-\mu_2^2\right]R=0.
\label{eq:AdS2-radial}
\end{equation}
Introduce
\begin{equation}
z=\frac{r-r_h}{r+r_h},\qquad \rho_0=-\frac{\mathrm{i}\omega}{2r_h},\qquad \nu:=\sqrt{\frac14+\mu_2^2},\qquad h_\pm:=\frac12\pm\nu,
\label{eq:AdS2-definitions}
\end{equation}
and
\begin{equation}
x:=-\frac{\mathrm{i}\omega}{r_h}.
\label{eq:AdS2-x}
\end{equation}
The hypergeometric parameters are then
\begin{equation}
a=h_-,\qquad b=x+h_-,\qquad c=1+x,\qquad d=2\nu.
\label{eq:AdS2-abc}
\end{equation}
The ingoing solution has the boundary expansion
\begin{equation}
{R_{\mathrm{in}}}\sim{C_{\mathrm{s}}}(1-z)^{h_-}+{C_{\mathrm{f}}}(1-z)^{h_+},
\label{eq:AdS2-asymptotics}
\end{equation}
with
\begin{align}
{C_{\mathrm{s}}}&=\frac{\Gamma(1+x)\Gamma(2\nu)}{\Gamma(x+h_+)\Gamma(h_+)},\notag\\
{C_{\mathrm{f}}}&=\frac{\Gamma(1+x)\Gamma(-2\nu)}{\Gamma(h_-)\Gamma(x+h_-)}.
\label{eq:AdS2-C}
\end{align}

Mixed AdS boundary conditions impose a linear relation between the two asymptotic coefficients \cite{KlebanovWitten1999,Witten2001,BerkoozSeverShomer2002,IshibashiWald2004}. On a finite-coupling chart we choose $(\alpha,\beta)=(\lambda,1)$, giving
\begin{equation}
F_\lambda(\omega)=\lambda{C_{\mathrm{s}}}(\omega)+{C_{\mathrm{f}}}(\omega)=0.
\label{eq:AdS2-F}
\end{equation}
Where the ratio is regular, this can be written as
\begin{equation}
\frac{{C_{\mathrm{f}}}}{{C_{\mathrm{s}}}}=\frac{\Gamma(-2\nu)\Gamma(h_+)\Gamma(x+h_+)}{\Gamma(2\nu)\Gamma(h_-)\Gamma(x+h_-)}=-\lambda.
\label{eq:AdS2-ratio}
\end{equation}
The projective limits $\lambda\to\infty$ and $\lambda=0$ select ${C_{\mathrm{s}}}=0$ and ${C_{\mathrm{f}}}=0$, respectively. Related Robin QNM spectra in asymptotically AdS black holes have been studied in Ref.~\cite{KinoshitaKozukaMurataSugawara2024}.

The numerical value of the Robin parameter depends on the normalization of the asymptotic basis. Since $1-z\sim2r_h/r$ as $r\to\infty$, write
\begin{equation}
R\sim A_{\rm s}r^{-h_-}+A_{\rm f}r^{-h_+},\qquad A_{\rm s}=(2r_h)^{h_-}{C_{\mathrm{s}}},\qquad A_{\rm f}=(2r_h)^{h_+}{C_{\mathrm{f}}}.
\label{eq:AdS2-physical-coefficients}
\end{equation}
If the boundary condition is expressed in physical coefficients as
\begin{equation}
A_{\rm f}+\lambda_{\rm phys}A_{\rm s}=0,
\label{eq:AdS2-physical-Robin}
\end{equation}
then
\begin{equation}
\lambda_{\rm phys}=(2r_h)^{2\nu}\lambda.
\label{eq:AdS2-lambda-physical}
\end{equation}
A parameter derivative must therefore specify whether $\lambda$ or $\lambda_{\rm phys}$ is held fixed when $r_h$ or $\nu$ is varied.

For real-frequency fields, the radial measure entering the Klein--Gordon norm is proportional to $\mathrm{d} r/(r^2-r_h^2)$. The slow branch behaves as $r^{-h_-}$, so its large-$r$ contribution scales as
\begin{equation}
\int^\infty r^{-3+2\nu}\,\mathrm{d} r.
\label{eq:AdS2-norm}
\end{equation}
Both falloffs are locally normalizable for
\begin{equation}
0<\nu<1,\qquad -\frac14<\mu_2^2<\frac34.
\label{eq:AdS2-window}
\end{equation}
Within the unregularized Kummer chart used in this paper we additionally require
\begin{equation}
2\nu\notin\mathbb Z.
\label{eq:AdS2-nonresonant}
\end{equation}
The latter is a restriction of the chosen hypergeometric basis, not of the admissible AdS boundary conditions themselves. The window \eqref{eq:AdS2-window} is the familiar range in which alternative or mixed scalar boundary data can arise \cite{BreitenlohnerFreedman1982,KlebanovWitten1999,IshibashiWald2004}.

The asymptotic Klein--Gordon flux is proportional to
\begin{equation}
\mathcal F_\infty\propto-2\nu\,\operatorname{Im}(A_{\rm s}^*A_{\rm f}).
\label{eq:AdS2-flux}
\end{equation}
Thus the condition \eqref{eq:AdS2-physical-Robin} has vanishing flux when $\lambda_{\rm phys}$ is real. Algebraically, complex Robin parameters may also be considered, but they do not define the same flux-conserving boundary problem. Vanishing flux by itself does not establish stability, positivity, or unitarity; those questions depend on the chosen extension \cite{IshibashiWald2004}.

The Green-function normalization follows from the same change of variables. With
\begin{equation}
p_2(r):=r^2-r_h^2,
\label{eq:AdS2-p}
\end{equation}
a unit radial source $\delta(r-r')$ transforms according to
\begin{equation}
p_2(r)\frac{\mathrm{d} z}{\mathrm{d} r}=2r_hz.
\label{eq:AdS2-weight-transform}
\end{equation}
After division by $\mathrm{d} z/\mathrm{d} r$, the delta-function Jacobian again leaves a unit source in $z$. Hence
\begin{equation}
w_{\rm AdS_2}(z)=2r_hz.
\label{eq:AdS2-weight}
\end{equation}
Since $w_0=z$ and $d=2\nu$,
\begin{equation}
D_{\rm AdS_2}=w_{\rm AdS_2}{W_{\mathrm{sf}}}=-4r_h\nu.
\label{eq:AdS2-D}
\end{equation}
At a simple mixed root on the finite-$\lambda$ chart, $\chi_n=1/{C_{\mathrm{s}}}(\omega_n)$, and the residue formula becomes
\begin{equation}
\operatorname*{Res}_{\omega=\omega_n}G(z,z';\omega)=-\frac{{R_{\mathrm{in}}}(z;\omega_n){R_{\mathrm{in}}}(z';\omega_n)}{D_{\rm AdS_2}\,{C_{\mathrm{s}}}(\omega_n)\,\partial_\omega F_\lambda(\omega_n)},
\label{eq:AdS2-residue}
\end{equation}
where
\begin{equation}
\partial_\omega F_\lambda=\lambda\,\partial_\omega{C_{\mathrm{s}}}+\partial_\omega{C_{\mathrm{f}}}
\label{eq:AdS2-Fprime}
\end{equation}
is taken at fixed background and fixed coordinate Robin parameter $\lambda$. If instead a physical coupling $\lambda_{\rm phys}$ is held fixed while background parameters vary, the relation \eqref{eq:AdS2-lambda-physical} must be differentiated as well. The spectral condition and the unit-source normalization are therefore kept distinct, as they will be in the P\"oschl--Teller calculation below.

\section{Pole coalescence in the P\"oschl--Teller/Nariai problem}\label{Sec:PT}

The P\"oschl--Teller potential provides one of the simplest settings in which the Green function, quasinormal spectrum, and pole coalescence can all be treated analytically. It has long been used in black-hole spectroscopy and appears in standard near-extremal de Sitter reductions \cite{PoschlTeller1933,FerrariMashhoon1984,CardosoLemos2003}. Related analytic treatments of black-hole QNMs beyond this exactly solvable model can be found in Refs.~\cite{ChurilovaKonoplyaZhidenko2022,KonoplyaZhidenko2023}. The critical higher-order-pole structure is also known from earlier analyses of the P\"oschl--Teller problem and, more recently, from the Nariai exceptional-line description \cite{Beyer1999,MaassenVanDenBrinkYoung2001,NakamotoOshita2026}. Our purpose here is narrower: with the source normalization fixed, we evaluate the finite-position simple-pole residues and both Laurent coefficients at the critical point. The full P\"oschl--Teller Green function has recently been analyzed in Ref.~\cite{Kuntz2026}; no new global Green-function construction is claimed here.

\subsection{Outgoing Green function and the critical point}\label{SubSec:PTCritical}

Consider
\begin{equation}
\frac{\mathrm{d}^2R}{\mathrm{d} x_*^2}+\left[\omega^2-\frac{V_0}{\cosh^2(\kappa x_*)}\right]R=0,\qquad \kappa>0,
\label{eq:PT-wave}
\end{equation}
for a perturbation channel whose Nariai or near-Nariai reduction takes this form. Introduce
\begin{equation}
z=\frac{1+\tanh(\kappa x_*)}{2}.
\label{eq:PT-z}
\end{equation}
The derivatives transform as
\begin{equation}
\partial_{x_*}=2\kappa z(1-z)\partial_z,\qquad \partial_{x_*}^2=4\kappa^2z^2(1-z)^2\partial_z^2+4\kappa^2z(1-z)(1-2z)\partial_z,
\label{eq:PT-derivatives}
\end{equation}
and Eq.~\eqref{eq:PT-wave} becomes
\begin{equation}
z(1-z)R''+(1-2z)R'+\left[\frac{\omega^2}{4\kappa^2z(1-z)}-\frac{V_0}{\kappa^2}\right]R=0.
\label{eq:PT-z-equation}
\end{equation}
The left-ingoing and right-outgoing endpoint factors are obtained with
\begin{equation}
\rho_0=\rho_1=-\frac{\mathrm{i}\omega}{2\kappa}.
\label{eq:PT-rho}
\end{equation}
After extracting $[z(1-z)]^{-\mathrm{i}\omega/(2\kappa)}$, the hypergeometric parameters satisfy
\begin{equation}
c=1-\frac{\mathrm{i}\omega}{\kappa},\qquad a+b=1-\frac{2\mathrm{i}\omega}{\kappa},\qquad ab=\left(-\frac{\mathrm{i}\omega}{\kappa}\right)\left(1-\frac{\mathrm{i}\omega}{\kappa}\right)+\frac{V_0}{\kappa^2}.
\label{eq:PT-abc-relations}
\end{equation}
Define
\begin{equation}
\epsilon:=1-\frac{4V_0}{\kappa^2},\qquad \eta:=\sqrt{\epsilon},\qquad u:=\frac12-\frac{\mathrm{i}\omega}{\kappa},
\label{eq:PT-epsilon}
\end{equation}
so that
\begin{equation}
a=u+\frac{\eta}{2},\qquad b=u-\frac{\eta}{2},\qquad c=u+\frac12.
\label{eq:PT-abc}
\end{equation}

At the right endpoint, the two basis solutions of Sec.~\ref{SubSec:Boundary} behave as
\begin{equation}
{R_{\mathrm{s}}}\sim e^{+\mathrm{i}\omega x_*},\qquad {R_{\mathrm{f}}}\sim e^{-\mathrm{i}\omega x_*},\qquad x_*\to+\infty.
\label{eq:PT-right-asymptotics}
\end{equation}
The outgoing condition therefore corresponds to $(\alpha,\beta)=(0,1)$, and the spectral function is
\begin{equation}
F(\omega,\epsilon)={C_{\mathrm{f}}}=\frac{\Gamma(c)\Gamma(c-1)}{\Gamma(a)\Gamma(b)}.
\label{eq:PT-F}
\end{equation}
Away from cancellations and singular parameter values, its zeros give the standard P\"oschl--Teller families \cite{Beyer1999,CardosoLemos2003},
\begin{equation}
\omega_n^\sigma=-\mathrm{i}\kappa\left(n+\frac12+\frac{\sigma\eta}{2}\right),\qquad \sigma=\pm1,\qquad n\in\mathbb{N}_0.
\label{eq:PT-spectrum}
\end{equation}

To fix the source convention, define the frequency-domain inverse by
\begin{equation}
\left[\partial_{x_*}^2+\omega^2-\frac{V_0}{\cosh^2(\kappa x_*)}\right]G_x(x_*,x_*';\omega)=\delta(x_*-x_*').
\label{eq:PT-G-definition}
\end{equation}
Set
\begin{equation}
\begin{aligned}
R_L(z;\omega,\epsilon)&=[z(1-z)]^{-\mathrm{i}\omega/(2\kappa)}{}_2F_1\!\left(a,b;c;z\right),\\
R_R(z;\omega,\epsilon)&:=R_L(1-z;\omega,\epsilon).
\end{aligned}
\label{eq:PT-RLR}
\end{equation}
Then $R_L$ is ingoing at the left horizon and $R_R$ is outgoing at the right horizon. Under the transformation \eqref{eq:PT-z}, the unit source in \eqref{eq:PT-G-definition} gives the divergence-form weight
\begin{equation}
w_x(z)=2\kappa z(1-z).
\label{eq:PT-weight}
\end{equation}
Since $d=c-a-b=\mathrm{i}\omega/\kappa$, Eq.~\eqref{eq:D} gives $D=-2\mathrm{i}\omega$, and the source-normalized kernel is
\begin{equation}
G_x(x_*,x_*';\omega)=\frac{R_L(z_<;\omega,\epsilon)R_R(z_>;\omega,\epsilon)}{2\mathrm{i}\omega\,F(\omega,\epsilon)}.
\label{eq:PT-G}
\end{equation}
No asymptotic approximation has been made at the source or observation point.

Although $a$ and $b$ contain $\sqrt{\epsilon}$, the physical quantities entering Eq.~\eqref{eq:PT-G} are analytic in $\epsilon$ near the critical point. In the hypergeometric series,
\begin{equation}
(a)_k(b)_k=\prod_{j=0}^{k-1}\left[(u+j)^2-\frac{\epsilon}{4}\right],
\label{eq:PT-Pochhammer-even}
\end{equation}
and $1/[\Gamma(a)\Gamma(b)]$ is likewise even in $\eta$. Thus the spectral function and Green-function numerator satisfy the local analyticity assumptions of Sec.~\ref{Sec:Coalescence}.

We now restrict the degeneracy statement to a positive barrier, $V_0>0$. Suppose $a(\omega_*)=-n$ is a zero of $F$ in the regular chart. Since $\partial_\omega a=-\mathrm{i}/\kappa\neq0$, the condition $F_\omega(\omega_*)=0$ requires the second reciprocal-Gamma factor to vanish as well, so $b(\omega_*)=-m$ for some $m\in\mathbb{N}_0$. Hence
\begin{equation}
\eta=a-b=m-n\in\mathbb Z.
\label{eq:PT-integer-difference}
\end{equation}
For $V_0>0$, $\eta$ is either real with $|\eta|<1$ or purely imaginary. The only integer possibility is therefore $\eta=0$. Conversely, at $\eta=0$ the two reciprocal-Gamma factors vanish linearly at the same frequency, while the numerator of Eq.~\eqref{eq:PT-F} is finite. Thus, within the present analytic sector,
\begin{equation}
V_0=\frac{\kappa^2}{4},\qquad a(\omega_{n,*})=b(\omega_{n,*})=-n,\qquad \omega_{n,*}=-\mathrm{i}\kappa\left(n+\frac12\right)
\label{eq:PT-critical-point}
\end{equation}
is precisely the positive-barrier locus at which the two same-overtone zeros of $F$ coalesce into a double zero. For fixed interior points with nonvanishing Green-function numerator, this double zero produces a second-order pole of the kernel. The critical P\"oschl--Teller/Nariai degeneracy itself is known \cite{Beyer1999,MaassenVanDenBrinkYoung2001,NakamotoOshita2026}; the argument above only fixes the range in which the discriminant criterion is valid.

The positive-barrier restriction is essential. For $\kappa=1$ and $V_0=-3/4$, one has $\eta=2$, and $\omega_*=-3\mathrm{i}/2$ gives $a=0$, $b=-2$, and $c=-1/2$. Direct expansion yields
\begin{equation}
F(\omega)=\frac{16\pi}{3}\left(\omega+\frac{3\mathrm{i}}{2}\right)^2+O\!\left[\left(\omega+\frac{3\mathrm{i}}{2}\right)^3\right].
\label{eq:PT-negative-well-counterexample}
\end{equation}
A negative well can therefore support a cross-overtone double zero at nonzero discriminant. At the separate free limit $V_0=0$, one has $a=c$ and $b=c-1$, so the Gamma factors cancel identically,
\begin{equation}
F(\omega,1)\equiv1.
\label{eq:PT-free-cancellation}
\end{equation}
These examples show why zeros of individual reciprocal-Gamma factors cannot be interpreted independently of the full spectral function.

\subsection{Finite-position residues and critical Laurent coefficients}\label{SubSec:PTLaurent}

Fix an overtone $n$ and define
\begin{equation}
q_n:=n+\frac12,\qquad c_n:=\frac12-n,\qquad \xi:=\sigma\eta.
\label{eq:PT-qcxi}
\end{equation}
At the pole $\omega_n^\sigma$, after exchanging $a$ and $b$ when necessary, the parameters can be written as
\begin{equation}
a=-n,\qquad b=-n-\xi,\qquad c=c_n-\frac{\xi}{2}.
\label{eq:PT-onshell-abc}
\end{equation}
The left solution then terminates after $n$ terms:
\begin{equation}
\phi_n(z;\xi)=[z(1-z)]^{-q_n/2-\xi/4}\sum_{k=0}^{n}\frac{(-n)_k(-n-\xi)_k}{(c_n-\xi/2)_k\,k!}z^k.
\label{eq:PT-phi}
\end{equation}
Up to normalization, this is the truncated P\"oschl--Teller mode appearing in the earlier completeness analysis of Ref.~\cite{Beyer1999}. Gamma reflection gives
\begin{equation}
{C_{\mathrm{s}}}(\omega_n^\sigma)=(-1)^n,
\label{eq:PT-Cs-parity}
\end{equation}
and hence
\begin{equation}
\phi_n(1-z;\xi)=(-1)^n\phi_n(z;\xi).
\label{eq:PT-parity}
\end{equation}

For $\eta\neq0$ sufficiently close to the critical point, the pole is simple. Using Eq.~\eqref{eq:rgamma},
\begin{equation}
F_\omega(\omega_n^\sigma)=-\frac{\mathrm{i}}{\kappa}(-1)^n n!\,\frac{\Gamma(c_n-\xi/2)\Gamma(-q_n-\xi/2)}{\Gamma(-n-\xi)}.
\label{eq:PT-Fomega}
\end{equation}
Substitution into Eq.~\eqref{eq:PT-G} gives the exact finite-position residue
\begin{equation}
\mathcal R_n^\sigma(z,z')=-\frac{\mathrm{i}\Gamma(-n-\xi)}{2n!\,\Gamma(c_n-\xi/2)^2}\,\phi_n(z;\xi)\phi_n(z';\xi),\qquad \xi=\sigma\eta.
\label{eq:PT-residue-exact}
\end{equation}
This expression retains the full spatial dependence at arbitrary fixed interior points and uses the unit $\delta(x_*-x_*')$ normalization of Eq.~\eqref{eq:PT-G-definition}.

To extract the finite term as the two poles coalesce, set
\begin{equation}
\phi_n(z):=\phi_n(z;0),\qquad v_n(z):=\left.\partial_\xi\phi_n(z;\xi)\right|_{\xi=0},\qquad f_z:=z(1-z),
\label{eq:PT-v-definition}
\end{equation}
and define
\begin{equation}
A_{nk}:=\frac{(-n)_k^2}{(c_n)_k\,k!},\qquad \mathcal H_m:=\sum_{j=1}^{m}\frac1j,\qquad \mathcal H_0:=0.
\label{eq:PT-Ank-H}
\end{equation}
Differentiating the terminating series gives
\begin{equation}
v_n(z)=f_z^{-q_n/2}\sum_{k=0}^{n}A_{nk}z^k\left[\mathcal H_n-\mathcal H_{n-k}+\frac12\sum_{j=0}^{k-1}\frac{1}{c_n+j}-\frac14\ln f_z\right],
\label{eq:PT-v}
\end{equation}
where the inner sum is zero for $k=0$. No Digamma function needs to be evaluated at a negative integer. This differentiation does not omit an off-shell tail: at $a=b=-n$, every term with $k>n$ contains two vanishing upper Pochhammer factors, so its first frequency derivative vanishes. Since $\epsilon=\xi^2$ along the two branches,
\begin{equation}
\left.\partial_\omega R_L(z;\omega,0)\right|_{\omega_{n,*}}=\frac{2\mathrm{i}}{\kappa}v_n(z).
\label{eq:PT-Romega-v}
\end{equation}

For later use, define
\begin{align}
P_n&:=(n!)^2\Gamma(c_n)^2=\frac{\pi16^n(n!)^4}{[(2n)!]^2},
\label{eq:PT-Pn}\\
d_n&:=\psi(c_n)-\psi(n+1)=2\left(\mathcal H_{2n}-\mathcal H_n-\ln2\right).
\label{eq:PT-dn}
\end{align}
Expanding Eq.~\eqref{eq:PT-residue-exact} at small $\eta$ gives
\begin{equation}
\begin{aligned}
\mathcal R_n^\sigma(z,z')={}&\frac{\sigma(-1)^n\mathrm{i}}{2P_n\eta}\phi_n(z)\phi_n(z')\\
&+\frac{(-1)^n\mathrm{i}}{2P_n}\left[d_n\phi_n(z)\phi_n(z')+v_n(z)\phi_n(z')+\phi_n(z)v_n(z')\right]+O(\eta).
\end{aligned}
\label{eq:PT-residue-expansion}
\end{equation}
The opposite $1/\eta$ divergences therefore have the same finite part. Since
\begin{equation}
\delta_n:=\frac{\omega_n^+-\omega_n^-}{2}=-\frac{\mathrm{i}\kappa\eta}{2},
\label{eq:PT-delta}
\end{equation}
comparison with Sec.~\ref{Sec:Coalescence} yields
\begin{equation}
K_{2,n}(z,z')=\frac{(-1)^n\kappa}{2P_n}\phi_n(z)\phi_n(z'),
\label{eq:PT-K2}
\end{equation}
and
\begin{equation}
K_{1,n}(z,z')=\frac{(-1)^n\mathrm{i}}{P_n}\left[d_n\phi_n(z)\phi_n(z')+v_n(z)\phi_n(z')+\phi_n(z)v_n(z')\right].
\label{eq:PT-K1}
\end{equation}
Together with Eqs.~\eqref{eq:PT-phi} and \eqref{eq:PT-v}, these are finite-sum expressions for both coefficients of the critical source-normalized Green kernel at arbitrary fixed interior points. The general Laurent identities are known \cite{MorikawaOgawaHirose2026}; the result here is their explicit evaluation in the P\"oschl--Teller problem, including the frequency derivative of the numerator.

As an independent check, define
\begin{equation}
\tau:=-\frac{\mathrm{i}(\omega-\omega_{n,*})}{\kappa},\qquad L_n:=2d_n+\frac1{q_n}.
\label{eq:PT-tauLn}
\end{equation}
Near $(\omega_{n,*},\epsilon=0)$, the spectral function has the joint expansion
\begin{equation}
F(\omega,\epsilon)=-\frac{P_n}{q_n}\left(\tau^2-\frac{\epsilon}{4}\right)(1+L_n\tau)+O(\tau^4,\tau^2\epsilon,\epsilon^2).
\label{eq:PT-F-local}
\end{equation}
In particular,
\begin{equation}
F_{\omega\omega,*}=\frac{2P_n}{q_n\kappa^2},\qquad F_{\omega\omega\omega,*}=-\frac{6\mathrm{i} P_nL_n}{q_n\kappa^3}.
\label{eq:PT-F-derivatives}
\end{equation}
For the numerator $H=R_L(z_<)R_R(z_>)/(2\mathrm{i}\omega)$,
\begin{equation}
H_*=\frac{(-1)^n}{2\kappa q_n}\phi_n(z)\phi_n(z'),
\label{eq:PT-Hstar}
\end{equation}
and
\begin{equation}
H_{\omega,*}=(-1)^n\left\{\frac{\mathrm{i}}{\kappa^2q_n}\left[v_n(z)\phi_n(z')+\phi_n(z)v_n(z')\right]-\frac{\mathrm{i}}{2\kappa^2q_n^2}\phi_n(z)\phi_n(z')\right\}.
\label{eq:PT-Homega}
\end{equation}
Substitution into Eqs.~\eqref{eq:K2-general} and \eqref{eq:K1-general} reproduces Eqs.~\eqref{eq:PT-K2} and \eqref{eq:PT-K1}. The numerator derivative is essential for the finite coefficient $K_{1,n}$.

With the inverse-frequency convention of Eq.~\eqref{eq:inverse-transform}, the isolated critical pair contributes
\begin{equation}
g_n^{\rm crit}(t;z,z')=e^{-\kappa q_nt}\left[-\mathrm{i} K_{1,n}(z,z')-tK_{2,n}(z,z')\right].
\label{eq:PT-time-critical}
\end{equation}
This is the standard second-order-pole time dependence \cite{MaassenVanDenBrinkYoung2001,NakamotoOshita2026}, now with the finite-position coefficients explicit. It is the contribution of the coalescing pole pair, not the complete causal Green function; other poles and non-pole contour contributions may also be required \cite{Kuntz2026,ArnaudoCarballoWithers2026}.

The parity relation \eqref{eq:PT-parity} also gives $v_n(1-z)=(-1)^n v_n(z)$. For odd $n$, both $\phi_n$ and $v_n$ vanish at $z=1/2$, so the entire pair principal part vanishes if either point lies at the center. At a generic node of $\phi_n$, only $K_{2,n}$ is forced to vanish. For the fundamental mode,
\begin{equation}
P_0=\pi,\qquad d_0=-2\ln2,\qquad \phi_0(z)=f_z^{-1/4},\qquad v_0(z)=-\frac14\ln(f_z)\phi_0(z).
\label{eq:PT-n0-data}
\end{equation}
At $x_*=x_*'=0$, equivalently $z=z'=1/2$,
\begin{equation}
K_{2,0}=\frac{\kappa}{\pi},\qquad K_{1,0}=-\frac{2\mathrm{i}\ln2}{\pi},\qquad g_0^{\rm crit}(t;0,0)=-\frac{\kappa t+2\ln2}{\pi}e^{-\kappa t/2}.
\label{eq:PT-n0-result}
\end{equation}
The overall sign follows from the convention in Eq.~\eqref{eq:PT-G-definition}, namely a unit $+\delta(x_*-x_*')$ source for the frequency-domain operator. Reversing the overall sign of the Green-function equation reverses all kernel coefficients together.

\section{Conclusions}\label{Sec:Conclusion}

We have used hypergeometric connection data to separate the spectral condition of a QNM problem from the normalization and spatial structure of its frequency-domain Green function. The BTZ and AdS$_2$ examples show how this separation works for standard and mixed boundary conditions. In the P\"oschl--Teller problem it leads to fully source-normalized, finite-position residues and to closed expressions for both coefficients of the critical Laurent expansion. In particular, the subleading coefficient retains the frequency dependence of the Green-function numerator; the derivative of the spectral function alone is not sufficient to determine the normalized response.

This result complements recent work on resonant excitation and exceptional points in black-hole spectra \cite{Motohashi2025,NakamotoOshita2026,PanossoMacedoKatagiriKubotaMotohashi2026,RossiOshitaGualtieriBerti2026} and the more general Riesz--Laurent description of sourced response near pole coalescence \cite{MorikawaOgawaHirose2026}. The exactly solvable P\"oschl--Teller model is useful here not because the existence of the double pole is new, but because the complete local coefficients can be evaluated analytically while keeping the source and spatial normalization explicit. The calculation also illustrates the need to work with the complete spectral and Green-function expressions rather than with individual Gamma-function factors, since cancellations can change the apparent pole structure.

Two extensions appear particularly natural. The first is the resonant hypergeometric sector, where integer exponent differences require logarithmic or regularized connection bases. The second is the extension to Heun-type connection problems relevant to rotating black holes, where recent calculations already show that excitation factors and exceptional structures remain important \cite{MotohashiSuichi2026,RossiOshitaGualtieriBerti2026}. A further step is to embed the local Laurent kernels into a complete causal decomposition, including source and observer projections and any non-pole contributions required by the full Green function \cite{Kuntz2026,ArnaudoCarballoWithers2026}. These questions provide a natural route from exactly solvable models to physically measurable ringdown response.

\section*{Statements and Declarations}

\paragraph{Funding}
No funds, grants, or other support was received.

\paragraph{Competing interests}
The author has no relevant financial or non-financial interests to disclose.

\paragraph{Author contributions}
Ye Zhou: Conceptualization, Methodology, Formal analysis, Investigation, Writing -- original draft, Writing -- review and editing.

\paragraph{Data availability}
No datasets were generated or analyzed during this study. All results are analytic and are contained in the manuscript.

\paragraph{Use of generative AI}
OpenAI ChatGPT was used during manuscript preparation for language editing, structural suggestions, drafting assistance, literature organization, and LaTeX formatting. All mathematical derivations, scientific claims, references, and final text were reviewed and verified by the author, who takes full responsibility for the content.

\bibliography{reference}

\end{document}